\documentclass[fleqn,usenatbib]{mnras}

\usepackage{newtxtext,newtxmath}
\usepackage[normalem]{ulem}

\usepackage[T1]{fontenc}
\usepackage{ae,aecompl}
\usepackage{comment}

\usepackage{graphicx}	
\usepackage{amsmath}	
\usepackage{bm}         
\usepackage[capitalize]{cleveref}

\newcommand{\ie}{\emph{i.e.} }

\renewcommand{\d}{{\rm d}}

\renewcommand{\vv}{\bm{v}}

\newcommand{\vnab}{\bm{\nabla}}

\newcommand{\subX}[2]{{#1}_\text{\tiny #2}}

\newcommand{\dl}{\subX{d}{L}}

\newcommand{\zlim}{\subX{z}{lim}}
\newcommand{\zcos}{\bar{z}}

\newcommand{\SGB}{{\rm SGB}}
\newcommand{\SGL}{{\rm SGL}}

\newcommand{\LCDM}{$\Lambda$CDM }

\newcommand{\Mpc}{{\rm Mpc}}
\newcommand{\Gpc}{{\rm Gpc}}
\newcommand{\kms}{\,{\rm km/s}}
\newcommand{\Mpch}{\,h^{-1}\,\Mpc}
\newcommand{\Gpch}{\,h^{-1}\,\Gpc}

\newcommand{\Msunh}{h^{-1}M_{\sun}}

\newcommand{\Hamlet}{{\scshape Hamlet}\ }

\usepackage{soul,xcolor}

\title[Bayesian reconstruction from peculiar velocities]{Testing Bayesian reconstruction methods from peculiar velocities}

\author[]{
	Aur\'elien Valade,$^{1, 2}$
	Noam I Libeskind,$^{1, 2}$  
    Yehuda Hoffman$^{3}$
    and Simon Pfeifer$^{1}$\\
	$^{1}$Leibniz-Institut für Astrophysik Potsdam (AIP), An der Sternwarte 16, 14482 Potsdam, Germany \\
	$^{2}$Univ Lyon, Univ Claude Bernard Lyon 1, CNRS, IP2I Lyon / IN2P3, IMR 5822, F-69622, Villeurbanne, France \\
	$^{3}$Racah Institute of Physics, Hebrew University, Jerusalem 91904, Israel
}

\date{Accepted XXX. Received YYY; in original form ZZZ}

\pubyear{2019}

\begin{document}

\label{firstpage}
\pagerange{\pageref{firstpage}--\pageref{lastpage}}
\maketitle

\begin{abstract}
    Reconstructing the large scale density and velocity fields from surveys of galaxy distances, is
    a major challenge for cosmography. The data is very noisy and sparse. Estimated distances, and
    thereby peculiar velocities, are strongly affected by the Malmquist-like lognormal bias. Two
    algorithms have been recently introduced to perform reconstructions from such data: the Bias
    Gaussian correction coupled with the Wiener filter (BGc/WF) and the \Hamlet implementation of
    the Hamiltonian Monte Carlo forward modelling. The two methods are tested here against mock
    catalogs that mimic the Cosmicflows-3 data. Specifically the reconstructed cosmography and
    moments of the velocity field (monopole, dipole) are examined. A comparison is made to the
    ``exact'' wiener filter as well - namely the Wiener Filter in the unrealistic case of zero
    observational errors. This is to understand the limits of the WF method. The following is found.
    In the nearby regime ($d \lesssim 40 \Mpch$) the two methods perform roughly equally well.
    \Hamlet does slightly better in the intermediate regime ($ 40 \lesssim d  \lesssim 120 \Mpch$).
    The main differences between the two appear in the most distant regime ($d \gtrsim 120 \Mpch$),
    close to the edge of the data. \Hamlet outperforms the BGc/WF in terms of better and tighter
    correlations, yet close to the edge of the data \Hamlet yields a slightly  biased
    reconstruction. Such biases are missing from the BGc/WF reconstruction. In sum, both methods
    perform well and create reliable reconstructions with significant differences apparent when
    details are examined. 
\end{abstract}




\section{Introduction}

\begin{figure*}
	\includegraphics[width=.329\textwidth]{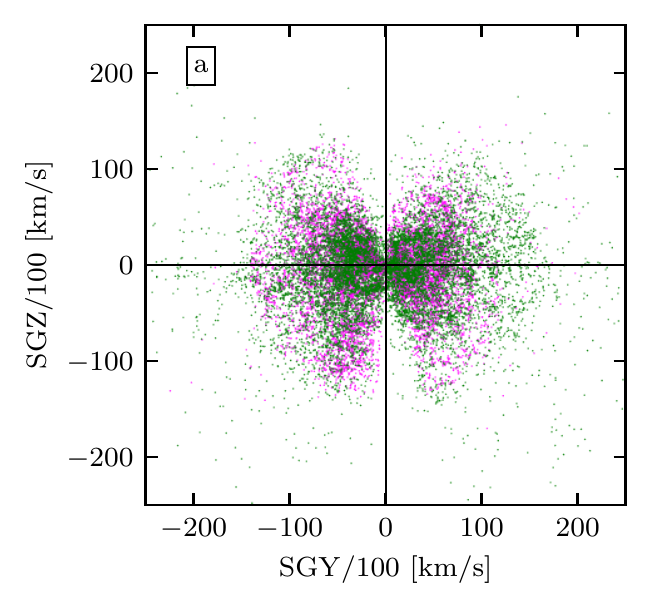}
	\includegraphics[width=.329\textwidth]{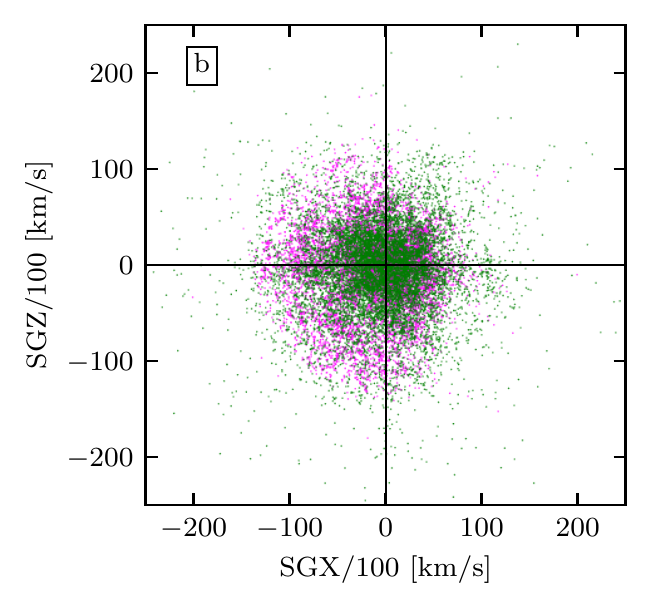}
	\includegraphics[width=.329\textwidth]{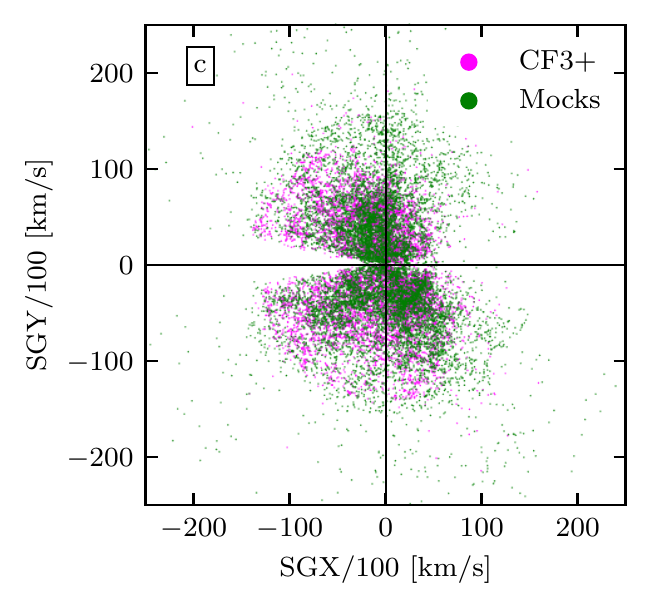}
    \caption{The distance, in units of km/s, of the CF3 data points (magenta) and mock data points
        (green) projected on the three supergalactic principal planes. Note the ZOA is accurately
    reproduced in the mock catalogues.}
	\label{fig:mocks_coordinates}
\end{figure*}

\begin{figure*}
	\centering
    \includegraphics[width=\textwidth]{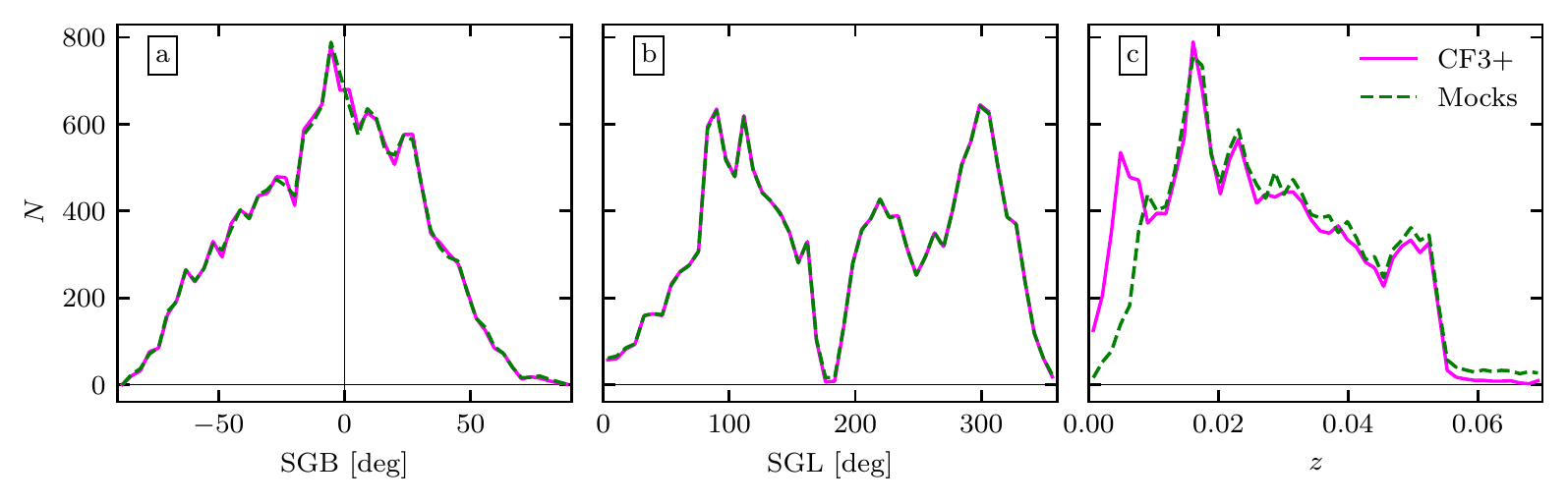}
    \caption{From left to right, the distribution SGB, SGL and $z$. Note that in panels a and b the
    two curves are on top of one an other.}
	\label{fig:mocks_distributions}
\end{figure*}

Mapping the large scale distribution of matter and its corresponding three-dimensional (3D) velocity
field is of great interest. The motivation is threefold. For one, mankind at large, and astronomers
in particular, explore their Universe by charting it. Making maps is a first step towards
understanding ones surroundings and ones place within a greater environment. Mapping the large scale
distribution of matter in the Universe is thus an end unto itself. Furthermore, explaining the large
scale velocity and density fields within the context of the \LCDM paradigm of structure formation
allows for the estimation of cosmological parameters that are inherent to that model
\citep{Jaffe1995, Pike2005, Feldman2010, Nusser2011, Carrick2015, Qin2019, Dupuy2019, Lilow2021,
Boruah2021a}. A third motivation is to map these fields at high redshift, thereby allowing for the
reconstruction of the initial conditions of the Local Universe \citep{Yepes2009, Sorce2016,
Hoffman2018, Libeskind2020, Sawala2021}.

We focus here on mapping the large scale structure (LSS) of the universe from surveys of peculiar
velocities, or rather from surveys of galaxies with measured distances from which the radial
component of the peculiar velocities may be extracted. Measuring galaxy distances is a formidable
challenge in observational cosmology. All distance measures rely on comparing an observed magnitude
with an (inferred or assumed) absolute magnitude. There are many ways to estimate a galaxy's
distance. For example, scaling relations tie the size of an elliptical galaxy or the angular
velocity of disc galaxy, to its intrinsic luminosity. Other methods include resolving stars at the
tip of the red giant branch, measuring Supernovae light curves, Cepheid variables pulsations or the
scale of fluctuations in a galaxies surface brightness. Each method has errors associated with it; a
combination of instrumentational errors, systematic errors, or calibration errors.

As such, compilations of peculiar velocities are difficult to analyze \citep[for a comprehensive
review][]{Straus1995} and are usually a patchwork of various surveys and methods observed with
different telescopes in different locations on earth (or in space). The POTENT method was the first
attempt to produce continuous maps of the density and velocity fields based on peculiar velocities
surveys \citep{Bertschinger1989}. The main underlying assumption of the POTENT method is that galaxy
velocities are drawn from an irrotational, potential flow. No further assumptions, were made on the
statistical nature of the flow field beyond the existence of a galaxy bias \citep{Kaiser1987}.
Therefore, its ability to handle the shortcomings of such
peculiar velocity surveys was limited. Subsequent approaches to the reconstruction of the LSS from
peculiar velocities have been formulated within Bayesian frameworks - these include the Wiener
filter (WF) and constrained realizations (CRs) methodology \citep{Ganon1993,Zaroubi1999,Tully2019}
as well as Monte Carlo Markov Chain algorithms \citep[MCMC][]{Lavaux2016, Graziani2019, Boruah2021b,
Prideaux2022}. They have been remarkably successful in ``mapping the invisible'' and recovering the
underlying cosmic fields.

Beyond the issues of noisy, sparse data, plagued with inhomogenous errors, there is one additional
inherent conceptual problem common to all surveys of peculiar velocities and that is that peculiar
velocity itself is not observed but is a \emph{derived} quantity. Given the redshift of and a
distance to a galaxy, it is the radial component of its peculiar velocity that can be computed. But
only the redshift is observed; distances themselves are not directly observed. What is measured is
the distance modulus of a galaxy \citep[cf.][]{Tully2016}. Because the error of the measured
distance modulus is assumed to be normally distributed, the errors on the observed distances are
thus log-normally distributed. This leads to a biased estimate of the distances and peculiar
velocities with respect to the actual distances \citep[see][]{Hoffman2021}. Often this  bias is
treated as yet another manifestation of the Malmquist bias \citep[see][]{Straus1995}. Here we refer
to it as the log-normal bias. For the WF/CRs reconstruction algorithm, the log-normal bias is
treated outside of the Bayesian framework in a separate process
\citep{Sorce2015,Tully2014,Hoffman2016,Hoffman2021}. For Monte Carlo methodologies the log-normal
bias is treated within a comprehensive algorithm \citep{Lavaux2016,Graziani2019}.

The Constrained Local UniversE Simulations (CLUES) project focuses on the reconstruction of LSS of
our nearby cosmic neighbourhood from surveys of galactic distances and thereby peculiar velocities,
in particular the Cosmicflows database \citep[cf.][and references therein]{Tully2016}. Two main
methodologies have been employed by the CLUES for the reconstruction of the local LSS and the
setting of  initial conditions for constrained cosmological simulations - one that is based on the
WF/CRs methodology \citep[][]{Hoffman1992, Zaroubi1995} and the other on MCMC and of Hamiltonian
Monte Carlo (HMC) sampling. In particular, within the WF/CRs framework the issue of the log-normal
bias has been handled by two independent algorithms, that of \citet{Sorce2015} and that of
\citet{Hoffman2021} and within the Monte Carlo sampling approach by \citet{Graziani2019} and
\citet{Valade2022}. 

Our aim in this work is to test the quality of two methods that reconstruct the LSS from peculiar
velocities: the WF/CRs method with a log-normal bias correction algorithm, known as the Bias
Gaussian Correction \citep[BGc;][]{Hoffman2021} and the HAmiltonian Monte carlo reconstruction of
the Local EnvironmenT (\Hamlet\ for short) method \citep{Valade2022}. These two methods are applied
to a mock data catalogue drawn from a cosmological simulation designed to imitate the CosmicFlows-3
data \citep{Tully2016}. The original simulation is referred to as the target simulation. The two
reconstructions are compared with the target simulation to gauge their fidelity. 

This paper is structured as follows. In \cref{sec:mocks} the algorithm for constructing halo
catalogues that mock the cosmic flows data is presented. In \cref{sec:BGc} the nature of the input
data as well as its biases and a bias correction scheme are presented. In \cref{sec:WF,sec:Hamlet}
the \Hamlet\ and WF/CR reconstructions methods are briefly described.  The results of applying to
the two reconstruction methods to these mocks, as well as a comparison between them is presented in
\cref{sec:results}. A summary and conclusion is offered in \cref{sec:conclusion}.

\section{Methods}

\subsection{Mock Catalogue construction}
\label{sec:mocks}

\begin{figure}
	\centering
	\includegraphics[width=\columnwidth]{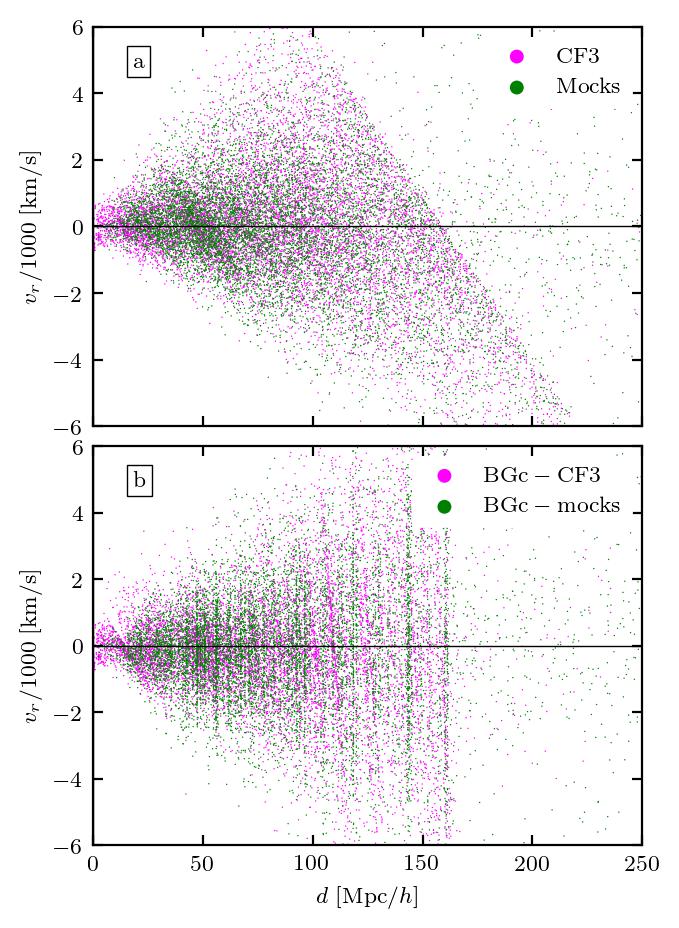}
    \caption{The distance of a galaxy as a function of its peculiar velocity is shown for the
        grouped CF3 data (magenta) as well as the mock catalogue (green). We can only make use of
        CF3 and not CF3+ as distances (and thus radial velocities) were not communicated for the
        pre-release of CF4. The log-normal bias is evident here in the lack of symmetry about
        $v_{r}=0$; beyond around 70Mpc$/h$ the universe appears to be systematically collapsing, in
        a so-called ``breathing mode''. Bottom panel: after application of the BGc correction, symmetry
    is reestablished.}
	\label{fig:mocks_vr_d}
\end{figure}

\begin{figure*}
	\centering
	\includegraphics[width=\textwidth]{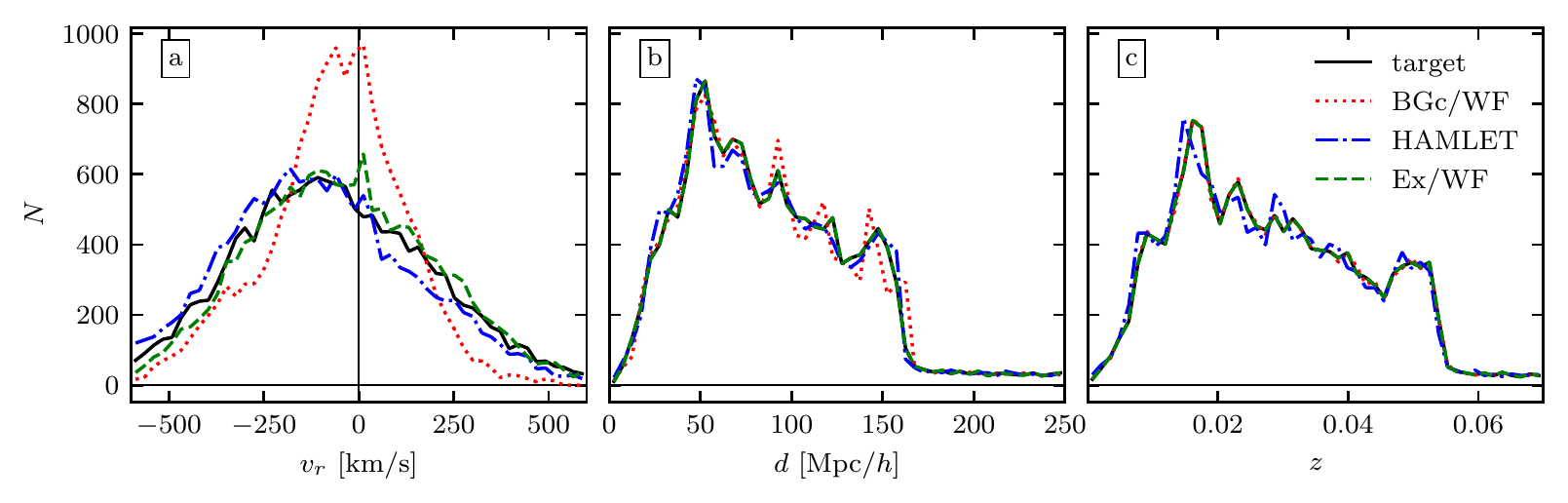}
    \caption{From left to right: a The distributions of the radial peculiar velocity, b the
        distance and c the redshift for the target (black solid lines), the BGc/WF (red dotted) and
    \Hamlet\ (Blue dash dotted) reconstruction methods. The Ex/WF method is shown in green dashed. } 
	\label{fig:mockspost_dist}
\end{figure*}

We wish to create a mock version of the grouped Cosmicflows catalogue that reproduces its main
characteristics, since it is on these observational data that the methods studied here will
eventually be applied (Valade et al, in prep). We start from the publicly available CF3 data release
and add to this $\sim4\,000$ points given to us by the authors of CF4 as a pre-release (Tully,
private communications)\footnote{R. B. Tully provided us with an advance set of redshifts and
angular positions of CF4 for the purpose of this paper. Distance modulii, and associated erros were
not provided.}, resulting in a ensemble of $\sim15\,000$ entries, hereafter named CF3+. 

A mock catalogue is constructed from the MultiDark Planck 2 simulation\footnote{The MultiDark
simulations are publicly available: \url{www.cosmosim.org}} \citep[MDPL2,][]{Riebe2013}, a dark
matter only $N$-body run of $N=3840^3$ particles in a periodic box of side length $L= 1 \Gpch$. The
cosmological parameters of the simulation are  from the 2nd Planck data release \citet{Ade2016} \ie\
a  flat \LCDM Universe $\Omega_{\rm m} = 0.307,~\Omega_{\rm b}=0.048,~ \Omega_{\Lambda}= 0.693,~
\sigma_{8} = 0.8228,~ n_{s} = 0.96$ and a dimensionless Hubble parameter $h= 0.678$ where $H_{0}=100
\times h\,\kms/\Mpc$ A Friend-Of-Friend's (FOF) algorithm with a linking length of 0.2 times the
mean inter particle separation is used to identify haloes whose mass is roughly $M_{200}$
\citep{Davis1985}. It is appropriate to use a FOF halo in this case since it is the \emph{grouped}
CF3 catalogue which is being mocked. Grouping the members of a virialized object together averages
out nonlinear motions implying that the (e.g.) cluster's peculiar velocity is a better traces of the
flow field. Note that the MDPL2 box size of $L= 1\Gpch$ is large enough to embed the CF3 catalogue,
whose effective depth is roughly $160\Mpch$.

An ``observer'' is associated with a randomly selected halo of mass in the range of $[0.9\,
\text{---}\,2.0]\times 10^{12}\Msunh$. The simulation is then re-centered on this halo and the
simulation's coordinate axes are then arbitrarily labelled as Supergalactic (SGX, SGY, SGZ).
Furthermore a mock ``sky projection'' is made such that each halo is given a sky position (SGL,
SGB).

The (proper) distance $d$ of each halo from the center, is used to compute a cosmological redshift
$\bar{z}$ by numerically integrating:
\begin{equation}
	\label{eq:intro:dl2zcos}
    d=c H{^{-1}_0} \int_0^{\bar{z}}\frac{1}{\sqrt{\Omega_m(1+\mathcal{Z})^3+(1-\Omega_m)}}\d{\mathcal{Z}}
\end{equation}
where $\Omega_m$ is the cosmological matter density parameter.  The (proper) distance $d$ is also
turned into a luminosity distances $\dl$ by
\begin{equation}
    \label{eq:intro:dl2d_vr}
	\dl=d  \times (1 + \zcos) 
\end{equation}
which is used to compute the halo's distance modulus:	
\begin{equation}
	\label{eq:intro:mu2dl}
    \mu = 5\log{\bigg(\frac{\dl}{\rm Mpc}\bigg)}+25  
\end{equation}

The radial peculiar velocity $v_{r}$ is combined with the cosmological redshift to obtain the full
redshift \citep{Davis2014}
\begin{equation}
    z+1=(\zcos+1)\left(\frac{v_{r}}{c}+1\right)
\end{equation}
At this point each halo's position relative to the observer has been transformed into two
``observable'' quantities: 1. a redshift $z$ (which includes a contribution from the radial
peculiar velocity $v_{r}$) and 2. a distance modulus $\mu$. 

The mock catalogue aims to have the same Probability Distribution Functions (PDF) of $P(\SGB)$,
$P(\SGL)$ and $P(z)$ as in the CF3+ data. This is
accomplished with a monte-carlo style algorithm in the following way: the same number of haloes as data points in CF3+ are drawn at
random from the simulation, within a sphere of around $300\Mpch$.  A merit is assigned to this
initial set of haloes by computing the absolute difference between its
$P(\SGB)$, $P(\SGL)$ and $P(z)$ and that of CF3+. Iterations proceed by adding and subtracting one
halo at a time and evaluating the merit of the new $P(\SGB)$, $P(\SGL)$ and $P(z)$, compared to
CF3+'s. If a new potential halo improves the merit of the distributions, it is kept; otherwise it is
rejected. In this way, the process converges halo by halo, towards reproducing the distribution
CF3+'s $P(\SGB)$, $P(\SGL)$ and $P(z)$. 

Once the merit function has converged and a suitable mock catalogue has been constructed, the
observational errors from the CF3 catalogue are added to the mock. Namely, the redshift and distance
modulus of each CF3 data point is given as $z+\varepsilon_{z}$ and $\mu+\varepsilon_{\mu}$ where
$\varepsilon_{z}$ and $\varepsilon_{\mu}$ denote the errors associated with each measurement.
$\varepsilon_{z}$ is assumed to be entirely due to spectroscopic precision while $\varepsilon_{\mu}$
depends on which standard candle is used and may range from 5\% for Supernova to 20\% for scaling
relations. Both $\varepsilon_{z}$ and $\varepsilon_{\mu}$ are assumed to be Gaussian with means of
zero and standard deviations of $c\sigma_{z}=50\kms$ and $ \sigma_{\mu}$, respectively. The value of
$\sigma_{\mu}$ associated to each halo is taken from the entry of CF3 whose redshift is the closest,
so as to reproduce the dependency of the $\sigma_{\mu}$ with the distance.

The fidelity of the mock to the CF3 catalogue, is shown in
\cref{fig:mocks_coordinates,fig:mocks_distributions}. \cref{fig:mocks_coordinates} shows the three
supergalactic projections with the mock data points in green and the CF3 constraints in purple. The
Zone of Avoidance (ZOA) and visual distribution of the catalogues are well recovered. Quantitatively
this is shown by looking at the distributions of $P(\SGB)$, $P(\SGL)$ and $P(z)$ themselves shown in
\cref{fig:mocks_distributions}a, b, and c, respectively. The distribution of SGB, SGL and $cz$ for
the mock galaxies and CF3 constraints, are largely indistinguishable from each other.
\cref{fig:mocks_distributions}c shows that within $\sim20\Mpch$, the number of CF3 constraints is
much greater than the mock catalogues presented here. This is because of there are too many CF3
constraints in this region, with respect to the resolution of our simulation. 

In principle the original unperturbed $d$, $d_{L}$, $\zcos$ and $v_{r}$ for each halo that in the
mock can be ``forgotten'' and new values can be computed using the values of $z$ and $\mu$ that
include the observational errors. These new values should exhibit similar biases to the
observational data by construction. This is seen in \cref{fig:mocks_vr_d}, where the radial velocity
as a function of distance is plotted. The diagonal cut in this plot is indicative of the log normal
bias dicussed in \cref{sec:BGc}. At a given distance there is an  unequal number of galaxies moving towards and
away from the observer, making it appear that the universe is contracting in a ``breathing mode''. This log normal bias and
its correction are presented in \cref{sec:BGc}.

\subsection{The log-normal bias and the Bias Gaussian correction (BGc)}
\label{sec:BGc}

One of the main purposes of constructing such a detailed mock catalogue as described above is to
ensure that the log-normal bias is reproduced, thereby allowing us to gauge the ability of the two
reconstruction methods to handle this bias. Much hand wringing and literature has been devoted to
the handling of biases in peculiar velocity surveys, and we refer the reader to \citet{Straus1995}
for a comprehensive explanation. Here we briefly explain what the log-normal bias is and how it is
handled in the context of the BGc as proposed by \citet{Hoffman2021}. We refer the reader to that
work for a comprehensive description of the log-normal bias and its correction by the BGc algorithm.

As mentioned, a Gaussian error on the distance modulus transforms into a log-normal error on the
luminosity distance (e.g. the inverse of \cref{eq:intro:mu2dl}). In other words, if the same galaxy
is observed many times, the mean of the different distance measures will not coincide with its
actual value. This bias changes the spatial distribution of the galaxies as well as their inferred
peculiar velocities. The log-normal bias can be seen in \cref{fig:mocks_vr_d} where the CF3 and mock
catalogue peculiar velocity $v_{r}$ is plotted as a function of distance. Beyond around
$\sim70\Mpch$, there is no longer symmetry in the distribution of $v_{r}$ about zero: more galaxies
have negative $v_{r}$ and the universe naively appears to be collapsing, a so-called ``breathing''
mode. In theory it can be corrected since the standard \LCDM\ model makes an explicit prediction
that the expected scatter for the radial component of the velocity is roughly $\sigma_{v}\sim 275
\kms$.

The essence of the BGc scheme is to map the log-normal distribution of the inferred distances
around their respective redshift distances into a normal distribution around the median of the
log-normal one. The width of that normal distribution is treated as a free parameter set to be $\sim
2 \Mpch$, in agreement with the \LCDM\ prediction that the intrinsic scatter of the radial
velocities is $\sigma_{v}\sim 275 \kms$. The same procedure is applied to the observed radial
velocities, retaining the median of the distribution of the radial velocities of data points in a
given redshift bin. Yet for the velocities, unlike the inferred distances, the variance of the
distribution is preserved as well. Namely the log-normal distribution of the observed distances is
mapped to a Gaussian distribution, while preserving the median of the log-normal distribution. It is
the invariance of the median under the normal - log-normal transformation which constitutes the
backbone of the BGc scheme. After the application of the BGc scheme to the data, the breathing mode
dissapears and the radial peculiar velocities scatter normally about 0 as can be seen in
\cref{fig:mocks_vr_d}, bottom panel.

\begin{figure}
	\centering
	\includegraphics[width=\columnwidth]{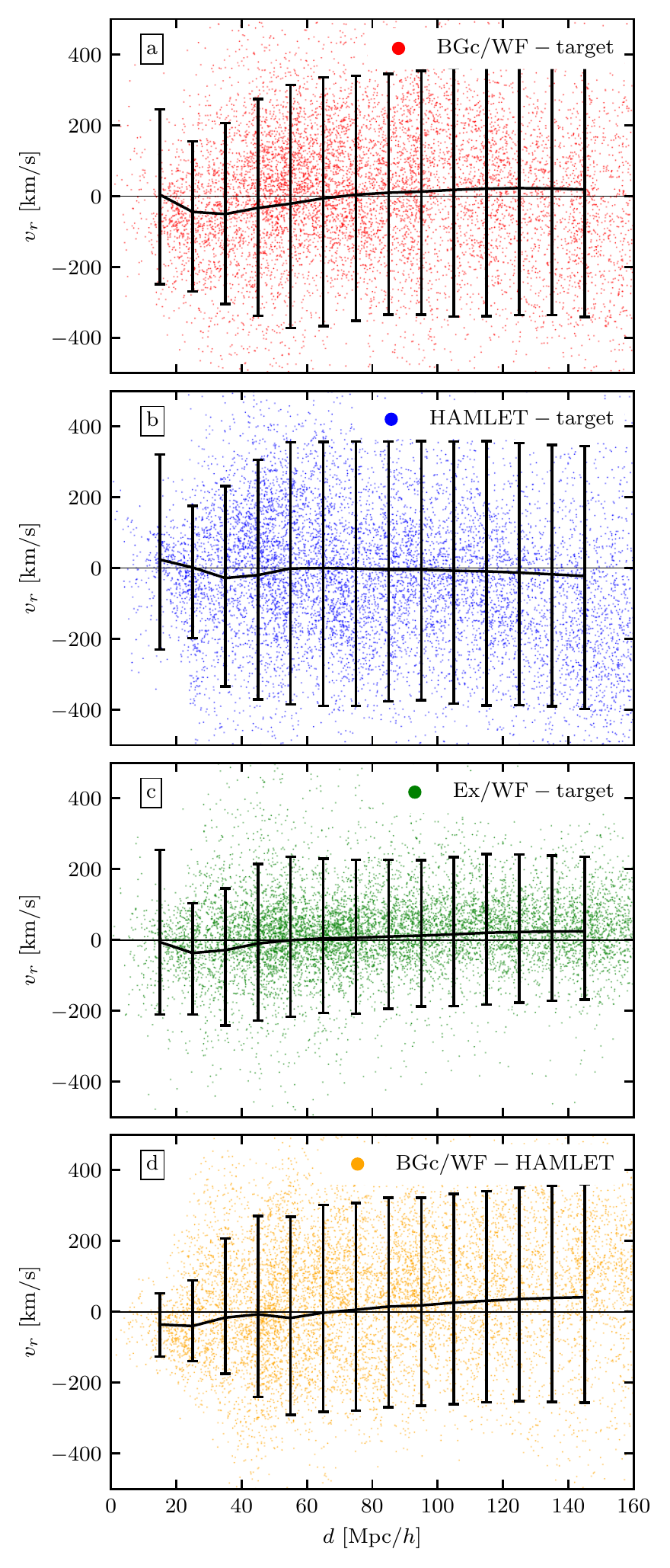}
    \caption{Scatter plots of the residual of the BGc/WF (panel a), of the \Hamlet\ (panel b) and
        of the WX/WF (panel c) reconstructed $v_{r}$s evaluated at the data points. The residual of the
    BGc/WF from the \Hamlet\ reconstructed $v_{r}$s at the data points is shown as well (panel d).  }
	\label{fig:mockspost_vr_d}
\end{figure}

\subsection{Wiener Filter and Constrained Realizations (WF/CRs)}
\label{sec:WF}

The WF/CR method is a tried and tested algorithm for reconstructing the large scale density
distribution of the universe, based on a limited number of peculiar velocity measurements. We refer
the reader to the voluminous literature \citep{Hoffman1992,Zaroubi1995,Zaroubi1999}
reviewing just the essentials here.

The WF is a Bayesian estimator of the underlying velocity field (and associated over-density) given a  set of observed radial velocities and a given assumed prior model of the
distribution of the peculiar velocities. In cosmological applications of the WF the \LCDM\
concordance cosmological model is taken to be the prior model. Accordingly: (a) the WF provides the
most probable continuous density and 3D velocity fields given a finite number of observed ``noisy''
radial velocities  and the assumed \LCDM\ model; (b) the CRs sample the constrained residual around
the WF field; and (c) the WF and CRs act so as to interpolate between data points and then
extrapolate beyond it. 

The WF/CRs methodology recovers the linear density and 3D velocity fields. In this context, the
over-density field and the 3D velocity are linked through the linearized coupled continuity and
Poisson equations 
\begin{equation}
    \delta =-\vnab\cdot\vv/H_0f(\Omega_m),
    \label{eq:delta}
\end{equation}
where $f(\Omega_m)$ is the linear growth factor. At large scales, or at early cosmological times,
the linear over-density is a good approximation for the fractional over-density,
$\delta=\rho/\bar{\rho}-1$, where $\rho$ is the density and $\bar{\rho}$ is the cosmological mean.
Unless stated otherwise, all the terms density and velocity refer to the \emph{linear} density and
velocity fields density. However, linear theory is clearly violated on small scales. Additionally,
the density field is more susceptible to non-linear dynamics than the velocity field. The linear
WF/CRs constitute a reasonable estimate of the actual velocity field down to the scale of roughly
$5-10 \Mpch$ \citep{Zaroubi1999}.

The WF estimator is the outcome of the ``tug-of-war'' between the data and the assumed prior, \LCDM\
in the present case. Where the data is has small errors, the estimated WF field is ``close'' to the
input data point. Otherwise, where the data is weak the WF solution is dominated by the prior model,
namely the solution tends to the mean density and null velocity fields. Consequently the constrained
variance, i.e variance spanned by the CRs, is small in the good data regime and converges towards
the cosmic variance as the data deteriorates.

\subsection{Wiener Filter reconstruction from Exact data (Ex/WF)}

As there exist no possibility to homogenize the sampling (namely the Zone of Avoidance will always
inhibit full sky coverage), the only source of uncertainty that could, one day, be mitigated, is the
observational uncertainty in the distance measurement. In order to test the methods' inherent
ability to reconstruct the underlying fields, an additional ``method'' is compared: the exact WF
(hereafter labeled Ex/WF). This is the WF applied to a mock where the error on each data point has
been artificially set to zero (and thus no BGc scheme is applied). This serves the purpose of
testing the WF in the case where the only source of uncertainty is the sampling. In other words in
\cref{sec:results}, the reconstruction based on the BGc/WF, the Ex/WF and the \Hamlet\ method are
presented.

\subsection{\Hamlet}
\label{sec:Hamlet}

The second approach examined here utilizes a Hamiltonian Monte Carlo sampling of the posterior PDF
via the HAmiltonian Monte carlo reconstruction of the Local EnvironmenT (\Hamlet\!\!) code, which is
described in full in \citet{Valade2022}.

Unlike the WF/CRs formalism, the \Hamlet\ algorithm treats the real distances of the data as unknown
dynamical variables that need to be estimated, much in the same way as the density and velocity
field. It is
Bayesian in nature because an \emph{ab initio} PDF of the variables that are to be estimated can be
specified.

A Monte-Carlo technique is used to sample the various posterior PDFs that are under consideration:
the cosmic density field, the velocity field and as the distances of the constraining data. The
technical challenges of the Monte Carlo approach is twofold. First the (assumed) prior  and
posterior  PDFs of the density and velocity field need to be extended to also include the
distribution of distances. Then an efficient sampling of the posterior PDF needs to be
devised. Given the extremely high dimensionality of the problem, the Hamiltonian Monte Carlo
algorithm is the tool of choice to perform that sampling, outperforming Metropolis Sampling or Gibbs
Sampling algorithms by many orders of magnitude \citep{Valade2022}.

Beyond its inhomogeneous distribution (e.g. the ZOA) the CF3+ catalogue has a fairly sharp cut off
at around a redshift distance  of $150\Mpch$ (see \cref{fig:mocks_distributions}c). In practice this
defines and limits the extent within which the \Hamlet\ reconstruction method is valid.
\citet{Hinton2017} investigated the problem that occurs when sampling from a distribution that has
an abrupt cut off. They show that the reconstructed fields (and inferred variables) close to the cut
off will be biased, effectively nullifying the applicability of the reconstruction beyond a small
amount closer than the cut off (say $\sim90\%$). It is therefore expected that any reconstruction
based on these data will not be valid beyond around $150\Mpch$.

\section{Results}
\label{sec:results}

\begin{figure*}
	\centering
	\includegraphics[width=\textwidth]{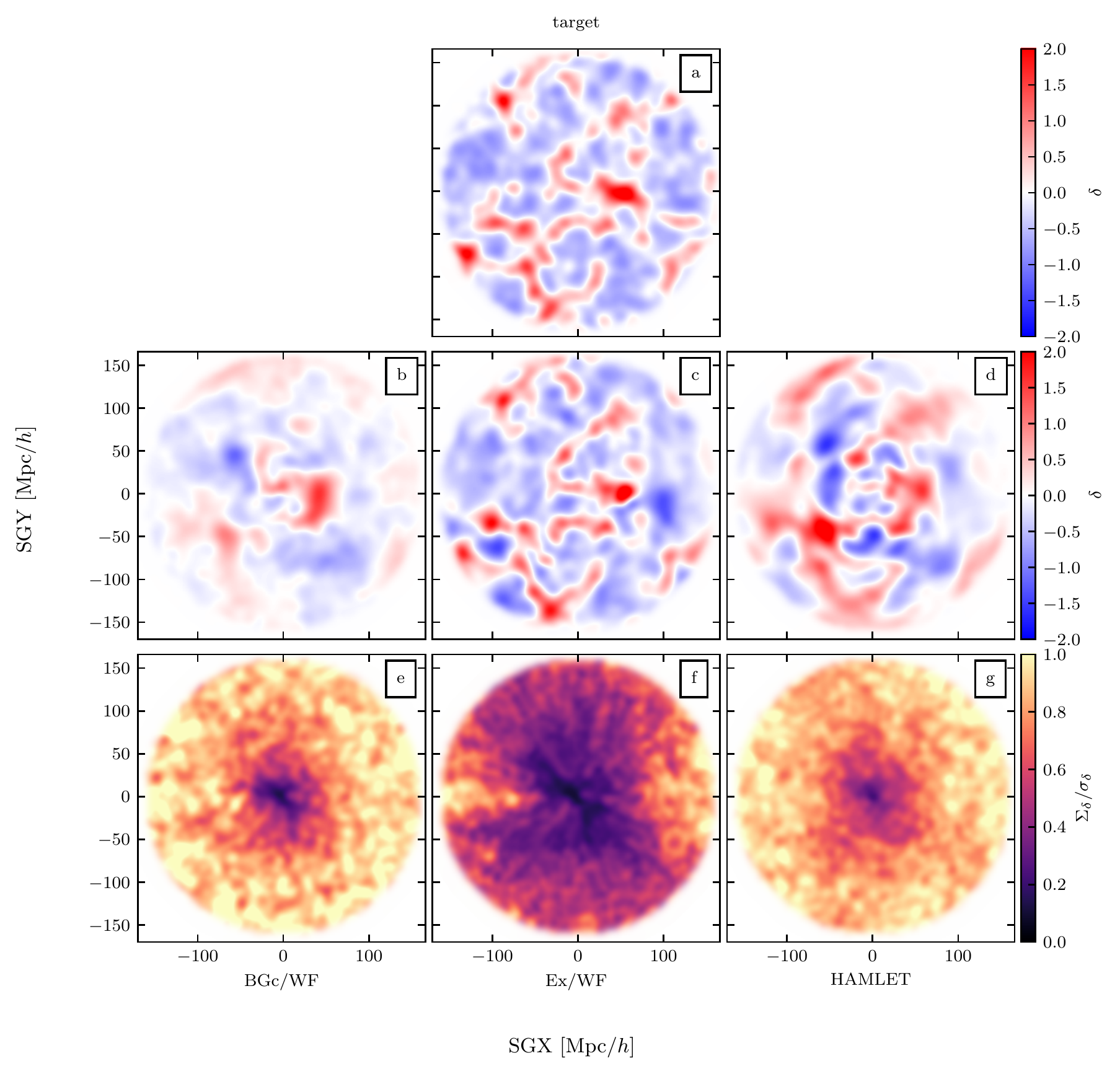}
    \caption{A comparison of the  of the BGc/WF (left column), Ex/WF (central column) and the
        \Hamlet\ (right column) reconstructed over-density fields with the target simulation.
        For consistency, the over-density plotted for the target field is the linear
        over-density \ie the divergence of the velocity field.  The middle panels present the
        reconstructed $\delta$ and the bottom ones show the constrained variance normalized by the
        cosmic variance, $\Sigma_\delta / \sigma_\delta$. All plots refer to the $SGZ=0$ plane of the
    target simulation and all fields are Gaussian smoothed with a $5\Mpch$ kernel.}
	\label{fig:divv_sgz}
\end{figure*}

\begin{figure*}
	\centering
	\includegraphics[width=1\textwidth]{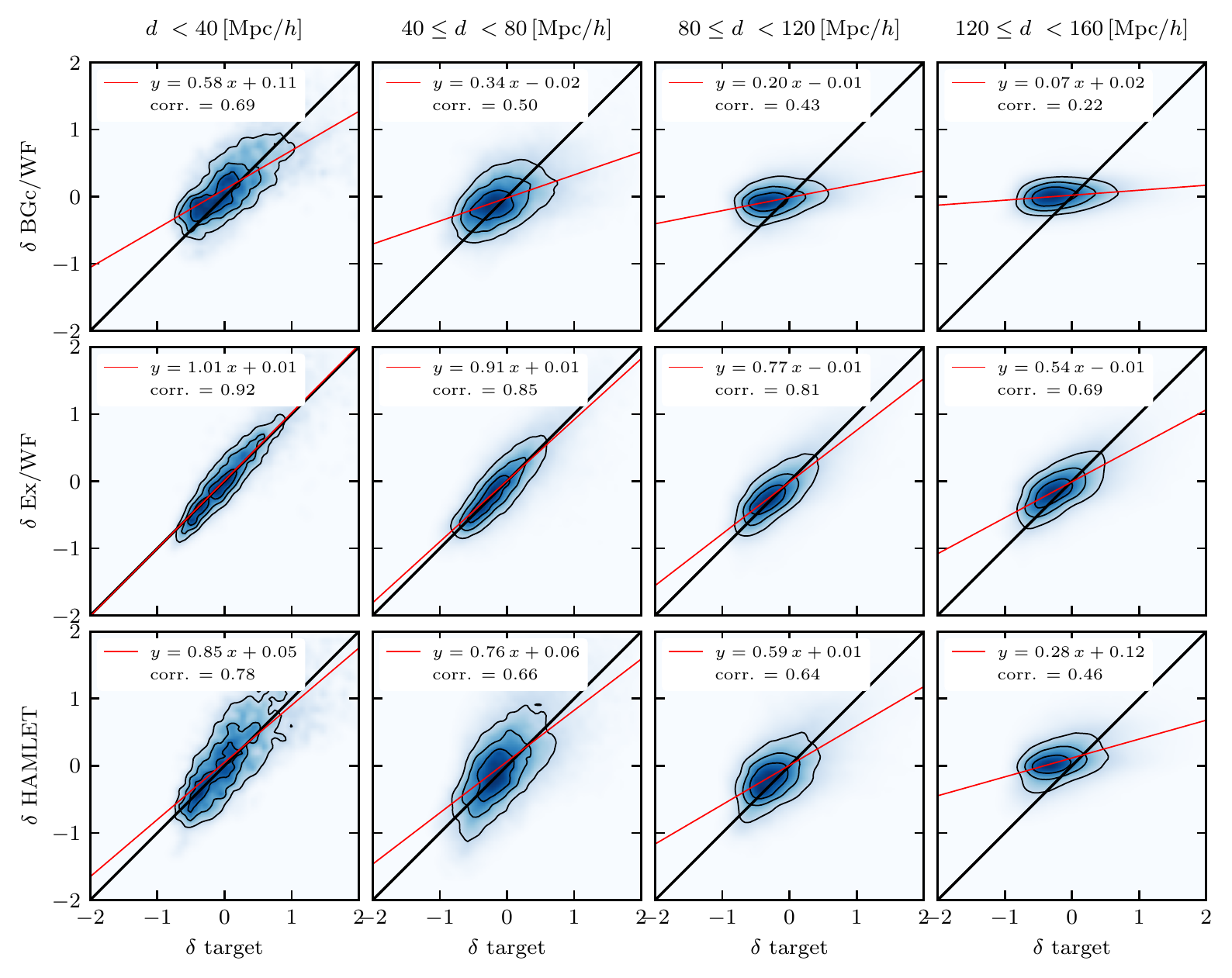} \\
    \caption{Density scatter plots of $\delta$ reconstructed versus $\delta$ target. Rows from
        bottom to top: \Hamlet, Ex/WF BGc/WF. Columns from left to right: within spheres of $40$,
        $80$, $120$, $160 \Mpch$. The red line represents the best fitted line whose line equation is
        $y=a x + b$. The parameters of the line and the Pearson correlation coefficient are given in
        the legend. The black line $ y = x $ is shown for reference.}
	\label{fig:scatters_divv}
\end{figure*}

\begin{figure}
	\centering
	\includegraphics[width=\columnwidth]{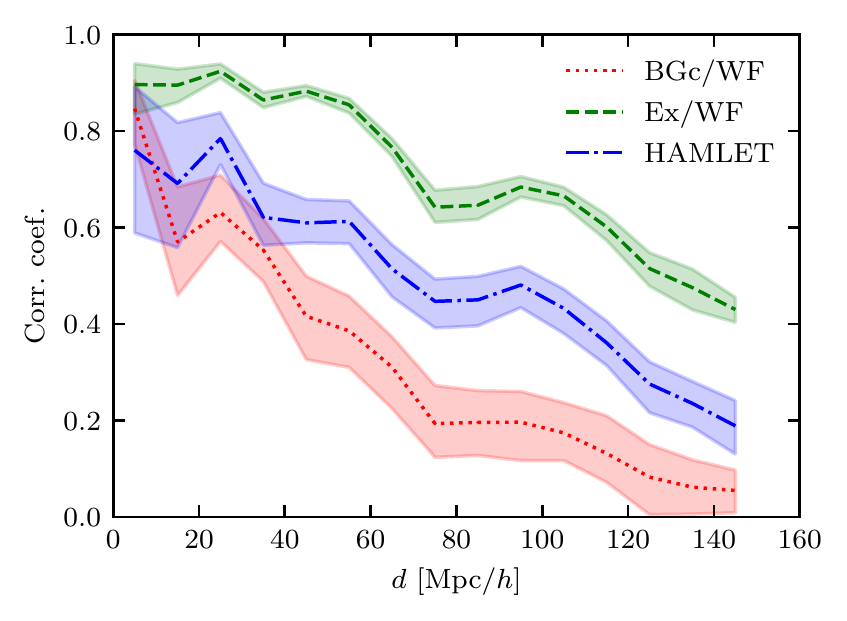}
    \caption{Statistics per shell of distance. Absolute value of the coefficient of correlation for
        $\delta$ between the different reconstructions and the target field. The error envelope
    represents the $2\sigma$ variation of the ensemble of realizations.}
	\label{fig:divv_corr}
\end{figure}

The results are presented in three sub-sections where we (a) compare how the predicted constraints
themselves differ from their real values (\cref{sec:data}); (b) examine the accuracy  of  the
reconstruction of the cosmic fields  (\cref{sec:maps}); and (c) compare the reconstructed monopole
and dipole (\ie bulk flow) multipoles with their target counterparts (\cref{sec:bulk}). 

\subsection{Reconstructed data}
\label{sec:data}

After applying the BGc/WF and \Hamlet\ methods to the mock catalogues (as well as the WF to the
exact, no error mocks), the first things to check is how well the distributions of radial peculiar
velocities, distances and redshifts of the data points, match their target values. This is shown
in \cref{fig:mockspost_dist}a,b and c, respectively, where the target curve represents the true
distributions of the mock catalogue; namely, the closer the BGc/WF or the \Hamlet\ curve is to the
target, the more accurate the reconstructions. The values of the reconstructed $v_{r}$'s of the mock data
points are obtained by interpolation over the grid points. The distances are obtained differently
for the different methods. The Ex/WF's distances are the true distances, thus they are identical to
the target. For the BGc/WF method, the distances are the result of the application of the BGc to the
data, before the WF is applied. Finally, for \Hamlet, the distance of each constraint is the mean of
all the distances sampled over the Monte-Carlo steps.

We remind the reader that the Exact WF (green dashed) represents the limits of the WF method.
\cref{fig:mockspost_dist}a shows that the \Hamlet\ reconstruction method does an exceptional
job at recovering the distribution of radial peculiar velocities. Note also that the WF in its
purest form too recovers the target distribution. The BGc/WF struggles slightly by narrowing the
data's distribution with a slight over emphasis on smaller values of the peculiar velocity at the
expense of the large values. We note, as an aside, that the fact that the target (and hence the
reconstructions) are not centered at $v_{r}=0$ is due to the specific nature of the mock observer
chosen (\ie cosmic variance). The BGc/WF suppression of the reconstructed radial velocities of the
data points relative to the target is inherent in the WF algorithm, where the estimated signal is
the weighted ``compromise'' between the  data and the prior model. Where the data is not very strong
the WF estimator is biased towards the null field predicted by the prior. 

In \cref{fig:mockspost_dist}b and c the distance and redshift distributions are examined.
For both of these quantities the two reconstructions do a remarkably good job at matching the
target, rendering their curves practically indistinguishable from the target. Note however that the
BGc/WF method tends to ``exaggerate'' some of the peaks and valleys in the distance distributions
(\cref{fig:mockspost_dist}b). All models reliably follow the input's form. 
In the absence of errors, \ie the Ex/WF case, the reconstructed $v_{r}$'s of
the  data points should be equal to the input constraints taken from the target simulation
\citep{Hoffman1992}.

The slight mismatch between the $v_{r}$ histograms of the Ex/WF and the target seen in
\cref{fig:mockspost_dist} occurs because the Ex/WF histogram is an interpolation over the course
grid of the WF.

It is important to understand by how much each constraint shifts during the BGc and reconstruction
procedures. In \cref{fig:mockspost_vr_d} the difference between the reconstructed $v_r$ and the
input $v_r$ is compared on a constraint by constraint basis and as function of distance. From top to
bottom this difference is shown for show the BGc/WF, \Hamlet and the Ex/WF (respectively
\cref{fig:mockspost_vr_d}a, b and c). The difference between the two main reconstruction methods
(BGc/WF and \Hamlet) is shown in the final panel, \cref{fig:mockspost_vr_d}d. In these plots each
constraint is a dot, the mean  value of the difference is shown as a black line and the standard
deviation of the distribution is designated with error bars. An examination of
\cref{fig:mockspost_vr_d} reveals that the methods based on the WF tend to underestimate the $v_r$
in the inner most distance shells (below $\leq 60\Mpch$) while overestimating it in the outer
shells. This is sure even for the ideal case of the Ex/WF. The mean of \Hamlet\ method, however
(\cref{fig:mockspost_vr_d}b) indicates the constraints are not systematically shifted in the region
$\sim 40 - 110 \Mpch$,  but underestimate $v_r$ outside this range.

\subsection{Reconstructed cosmic fields}
\label{sec:maps}
In this section the reconstructed density and velocity fields are examined and compared with the target.

\begin{figure*}
	\centering
	\includegraphics[width=\textwidth]{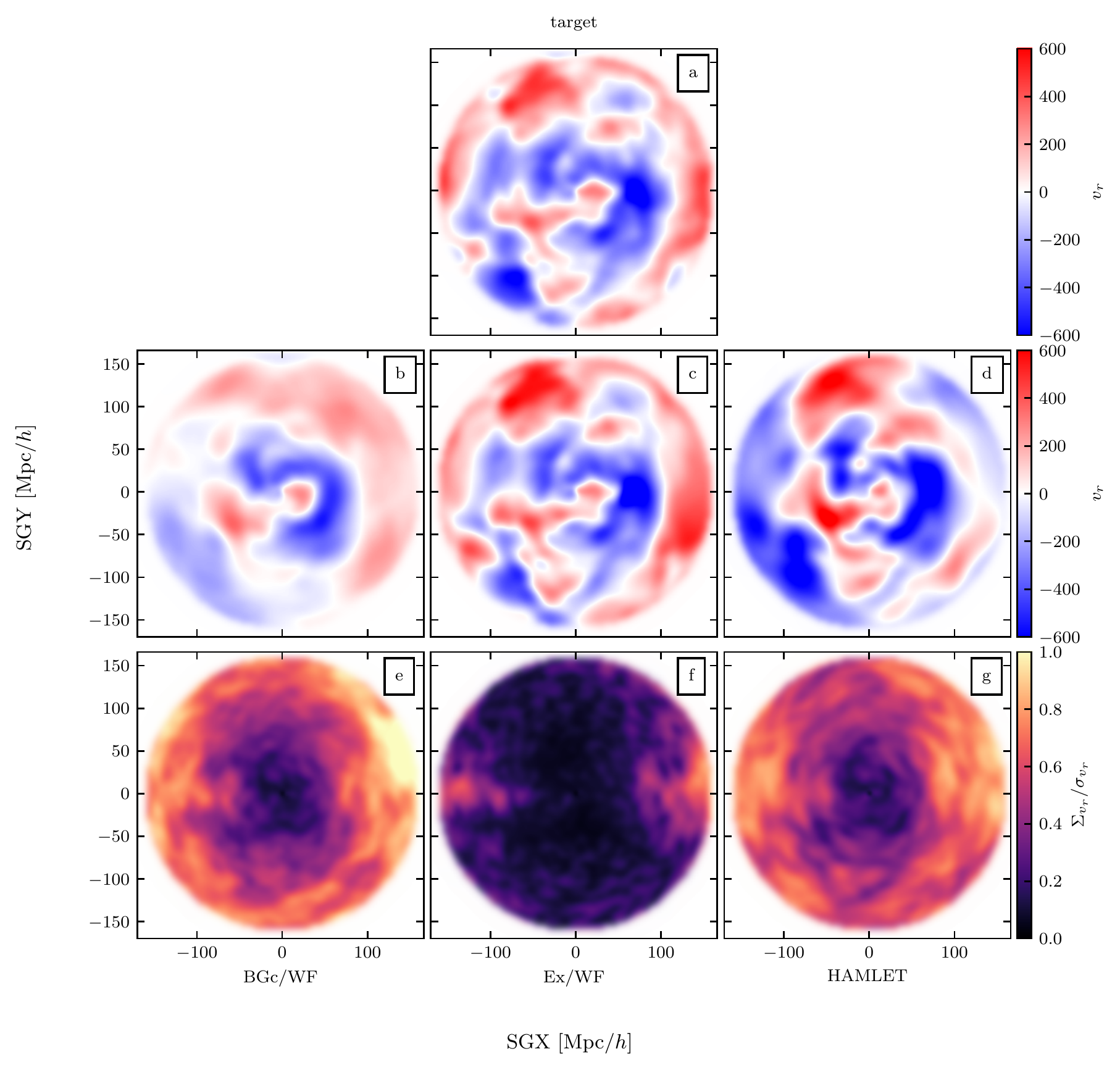}
    \caption{Same as \cref{fig:divv_sgz} for the radial component of the velocity field.}
	\label{fig:vr_sgz}
\end{figure*}

\begin{figure*}
	\centering
	\includegraphics[width=1\textwidth]{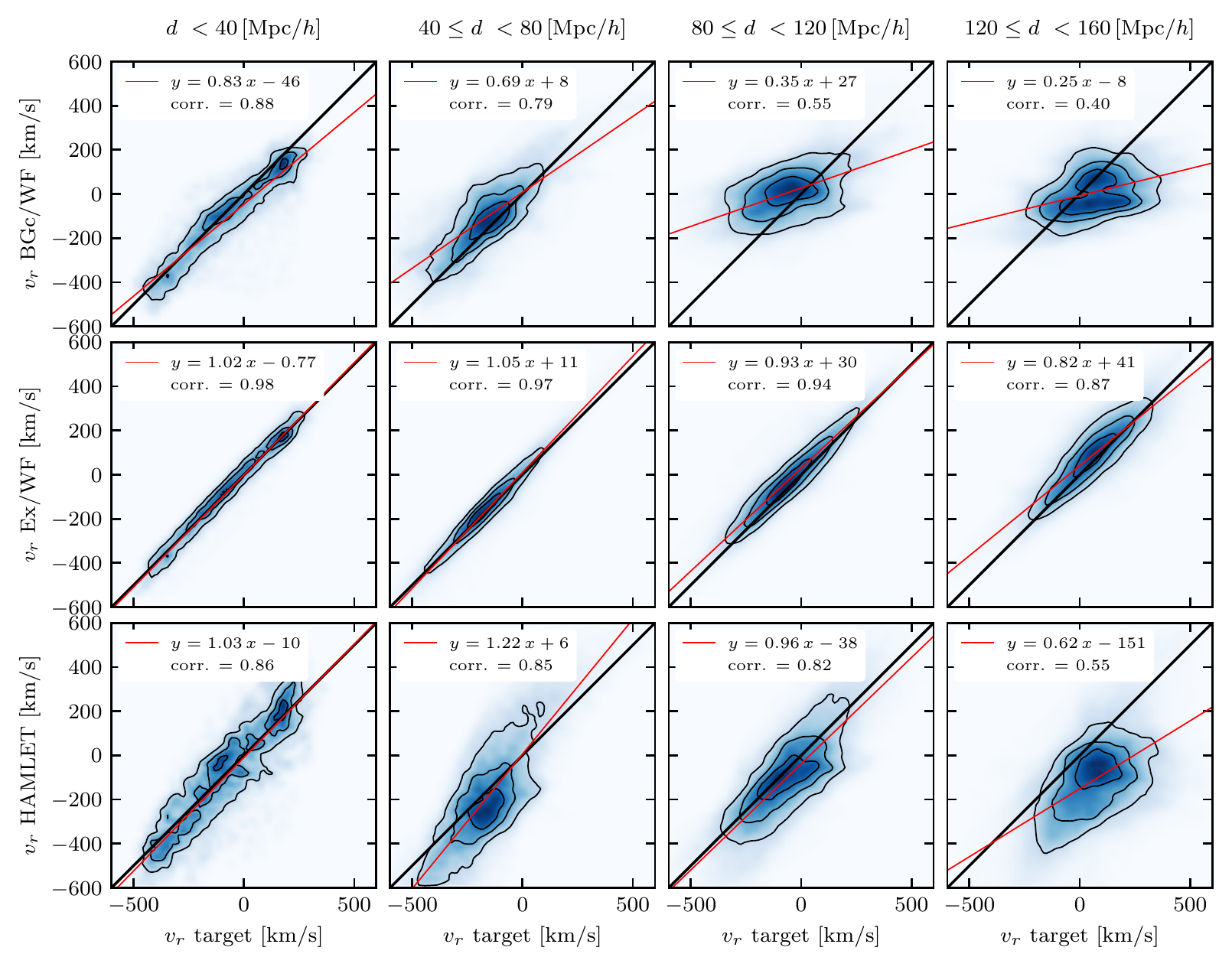} \\
    \caption{Same as \cref{fig:scatters_divv} for the radial component of the velocity field.}
	\label{fig:scatters_vr}
\end{figure*}

\begin{figure}
	\centering
	\includegraphics[width=\columnwidth]{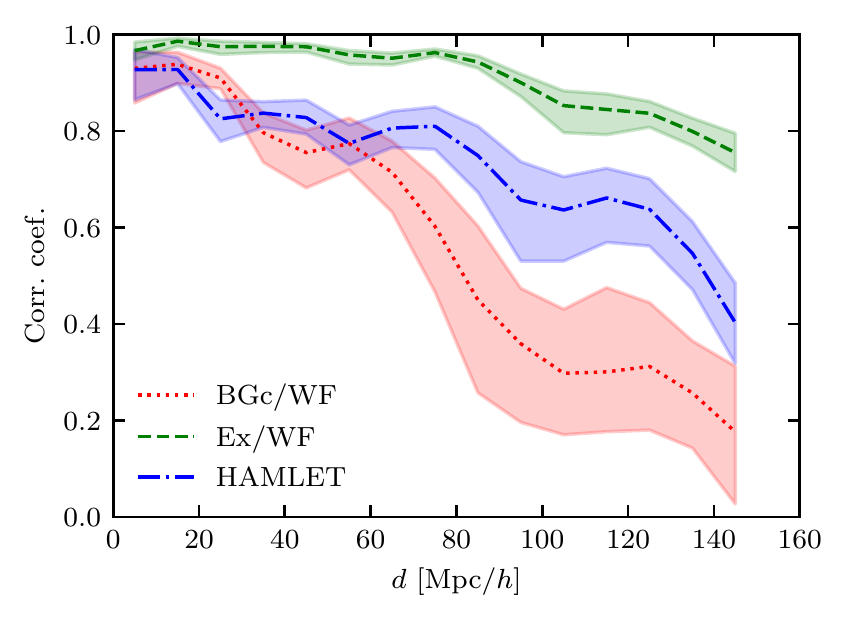}
    \caption{Same as \cref{fig:divv_corr} for the radial component of the velocity field.}
	\label{fig:vr_corr}
\end{figure}

\begin{figure}
	\centering
	\includegraphics[width=\columnwidth]{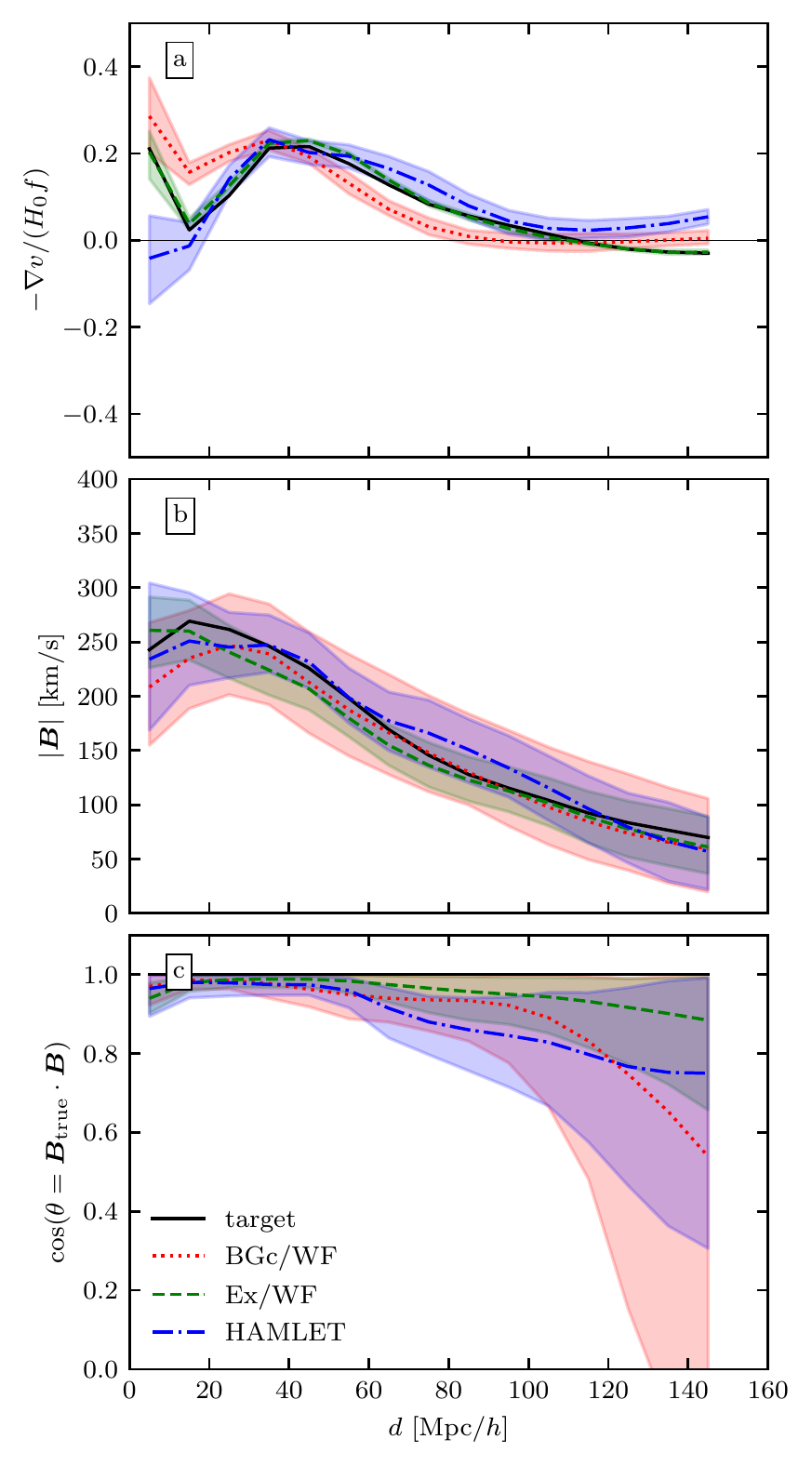}
    \caption{The monopole moment (upper panel), the amplitude of the dipole moment (\ie the
            bulk velocity; middle panel), and the cosine of the angle of alignment between the
            reconstructed and target bulk velocities (lower panel) are shown. The profiles present
            the mean and ``$2 \sigma$'' scatter of the mean profile in spheres of radius $d$. The
            reconstructions correspond here to the Ex/WF (green dashed line), the BGc/WF (red dotted
            line) and the \Hamlet\ (blue dot-dashed line) case. The scatter is the constrained
            variance of the different reconstruction and the target  simulation is presented by the
            black solid line (middle and upper panels).}
	\label{fig:moments}
\end{figure}

\subsubsection{Reconstructed density maps}

The non-linear density field of the target simulation cannot be directly compared with the
reconstructed linear density field. To enable a meaningful comparison we compare the divergence of
the velocity field of the two (\ie \cref{eq:delta}), terming both of these $\delta$, out of
convenience.

\label{sec:recon_dens}

In order to visually inspect the reconstructed density distribution, a $3.9\Mpch$ thick slab at the
super galactic plane (SGZ=0) is chosen. This is not an arbitrary choice: given that the largest
numbers of constraints are expected to lie in or close to SGZ=0, we expect this slab to be the most
accurate. The fields are smoothed with a Gaussian kernel of $5\Mpch$. 

\cref{fig:divv_sgz} examines the density distribution in this slab.  \cref{fig:divv_sgz}a is the
target density distribution. The column below it (namely the middle column, \cref{fig:divv_sgz}c, f
shows the Ex/WF results, while the left column (\cref{fig:divv_sgz}b, e) shows the BGc/WF results
and the right column (\cref{fig:divv_sgz}d, g) shows the \Hamlet\ results. The middle row (panels b,
c, d) shows the reconstructed density distribution. Some conclusions may be drawn from a visual
examination of \cref{fig:divv_sgz}b, c, d. The Ex/WF generally recovers the features of the local
cosmography at all distances. The reconstruction is not exact; given that there are no
``observational'' errors here, this implies that the mismatch between the Ex/WF and the target (\ie
between \cref{fig:divv_sgz}c and \cref{fig:divv_sgz}a), are entirely due to the finite, inhomogenous
and anisotropic sampling. Comparing the BGc/WF (\cref{fig:divv_sgz}b) with the target indicates a
decline of power of the reconstructed density field with the distance from the observer, yet the
general structure of the cosmic web of over- and under-dense regions is recovered. The \Hamlet
reconstructed $\delta$ field does not exhibit the same loss of power as in the BGc/WF case but it
suffers from a loss of spatial resolution with distance (\cref{fig:divv_sgz}d). The more distant
structures become more fuzzy and diffuse. 

The bottom panels of  \cref{fig:divv_sgz} present the constrained variance $\Sigma^2_{\delta}$ of
the three reconstructed $\delta$ fields. It is defined as the local, cell by cell, variance
calculated over an ensemble of CRs for the Ex/WF and BGc/WF case and over a set of independent
states of the Monte Carlo chain in the \Hamlet case. The panels show the square root of constrained
variance normalized by the cosmic variance, $\Sigma_{\delta}/\sigma_{\delta}$. The cosmic variance
is calculated by calculating the variance over all CIC cells in each reconstructed $\delta$ field.
The value of $\Sigma_{\delta}/\sigma_{\delta}$ gauges the constraining power of the constraints and
the assumed prior model. When this equals to 0 the region is highly constrained and when it equals
unity the reconstructions are as random as cosmic variance. Thus one expects it to be small close to
the observer and to approach unity asymptotically with distance. 

\cref{fig:divv_sgz}e, f, g quantifies what is visually apparent from (\cref{fig:divv_sgz}b, c, d)
namely that the inner regions are well constrained but that this fades with increasing distance. The
reconstruction methods that include errors (\ie \cref{fig:divv_sgz}e,g) are never ``perfect'', while
the Ex/WF method \cref{fig:divv_sgz}f, does obtain values of $\Sigma_{\delta}/\sigma_{\delta}$
close to 0. Interestingly, the impact of the ZOA on the reconstruction method is apparent in
\cref{fig:divv_sgz}f. Here it causes a very clear limitation of the expected ability
to reconstruct the density field.

The accuracy of the density field reconstructions - specifically their accuracy \emph{as a function
of distance} - is shown in \cref{fig:scatters_divv}. These are scatter plots which compare, on a
cell by cell basis, the density of the target with the BGc/WF (top row), Ex/WF (middle row) and
\Hamlet (bottom row). The line, $y=a x + b$, (or $\delta_{\rm method} = a \delta_{\rm target} +b$, to be more precise), which best describes the scatter
is shown in red; its slope, $y$ intercept and the Pearson correlation coefficient is given in each sub-panel. In
the ideal case where a reconstruction method perfectly matches the target this would simply be a
slope of $a=1$ and an offset or bias of $b=0$ line with zero scatter (shown in black), with a
Pearson correlation coefficient of unity. The columns in this figure denote different $40\Mpch$
thick radial shells under consideration. Note that a slope less than unity indicates that the reconstruction under-estimates the
over-dense regions and over estimates the under-dense regions. A slope greater than unity represents
the opposite (exaggerates over- and under-dense regions). An offset of $b\neq0$ means a biased
reconstruction.

There are a number of important features of this \cref{fig:scatters_divv}. First, considering the
inner most bin (leftmost column) the Ex/WF reconstruction recovers very well the density of the
target. A slope of unity  and practically null offset of $b=0.01$ and a correlation coefficient of
0.92 indicates that in general in this region the Ex/WF reconstructions is  very well recovered.
This implies that the nearby sampling of the CF3 catalog is almost optimal. Obviously, the \Hamlet
and the BGc/WF methods do worse in recovering the density field. Moving to the outer shells all
three density reconstruction systematically degrade with slopes and correlation coefficient
decreasing. The slope in all cases is less than unity, indicating that the reconstructions suppress
the power of the recovered density field. This diminishing of the power increases with the distance
from the observer. The BGc/WF suffers more from the loss of correlation with distance than the
\Hamlet. Yet, the latter reconstruction is at large distances with $b=0.12$ for the distance range
of $120\ \leq\ d\ \leq\ 160\ Mpch$. The BGC/WF behaves, on the other hand, by the ``Bayesian book''
- where the sampling is very sparse and the errors are much larger than the signal, the unbiased
\LCDM prior is recovered. 

The correlation between the reconstructed mean field and the target is shown as a function of
distance in \cref{fig:divv_corr}. These are the correlation coefficients from the scatter plots
(\cref{fig:scatters_divv}) plotted as a function of distance in order to gauge the degradation of
the reconstruction methods as data becomes more sparse and volumes become large. Note that the binning is
different hence the non identical
values of the correlation coefficient between the two plots. The solid lines in \cref{fig:divv_corr}
represent the mean correlation coefficient between the reconstruction and the target; the error
corridor represents the 2$\sigma$ variance about this mean. As expected the Ex/WF is always a
superior to the BGc/WF and the \Hamlet method. With the exception of the inner most bin, the
\Hamlet\  method achieved higher correlation coefficients than the BGc/WF method. At the edge of the
data, no method achieves a correlation coefficient of greater than 0.5.

\subsection{Reconstructed radial velocity maps}
\label{sec:bulk}

The examination of the radial component of the velocity field follows here that of the density field
(\S \ref{sec:recon_dens}). The same $SGZ=0$ and $4\Mpch$ thick slab is shown in \cref{fig:vr_sgz}.
Again, the top panel is the target radial peculiar velocity field while the left column shows the
BGc/WF reconstruction, the middle column the Ex/WF reconstruction and the right most column, the
\Hamlet\ reconstruction. The fields are smoothed with a Gaussian kernel of $5\Mpch$. 

The radial velocity field (\cref{fig:vr_sgz}b, c, d) appears much more accurately reconstructed
than the density field. The same outflows and inflows are generally visible and the cosmographic
landscape is recognisable in all three cases. Although the reader will note that the accuracy of the
velocity reconstruction, like the density field, deteriorates at larger distances. Features are
recognisable but distorted and smoothed out.

The constrained and cosmic variances of the radial velocity, $\Sigma_{v_{r}}$ and $\sigma_{v_{r}}$,
are calculated much in the same way as in for the density field (\S \ref{sec:recon_dens}). The
imprint of the ZOA is clearly seen in the $\Sigma_{v_{r}}/\sigma_{v_{r}}$ map of the Ex/WF map. Yet,
in all cases considered here the constrained variance, normalized by the cosmic variance, is much
smaller than in the density case. Namely, the velocity field is much more constrained by the CF3 data
than the density field. In general the reconstructed \Hamlet\ velocity field bares a closer resemblance to the target than the BGc/WF reconstruction.

A close inspection of \cref{fig:vr_sgz}d uncovers one troubling feature. At the edge of the
reconstructed volume, at distances close to $150 \Mpch$ the reconstructed is ``bluer'' than in the
corresponding target and the Ex/WF maps. Namely the \Hamlet\ reconstructed velocity field has a
spurious negative infall. This is a manifestation of the limitation of the method as described by \citet{Hinton2017} in \cref{sec:Hamlet}.

Again, we turn to a scatter plot, on a cell by cell basis to quantify the quality of the
reconstruction as a function of distance in \cref{fig:scatters_vr}, which is structured identically
to \cref{fig:scatters_divv} - namely radial extent increasing column wise from left to right, while
the rows from top to bottom being BGc/WF, Ex/WF and \Hamlet. This figure is qualitatively identical
to its density field counter part (\cref{fig:scatters_divv}) in that the same behavioural trends
between the different reconstructions methods and as a function of distance exist.  The correlation
analysis of the Ex/WF and BGc/WF cases behaves much in the same way as for the density field - a
degradation of the correlation with distance, a slope ($a$) that is close to unity nearby and
diminishes with distance, and essentially  with zero offset ($b\sim 0$). Yet, the quality of the
reconstruction of the radial velocity is much better than that of the density. The \Hamlet\
reconstruction shows a somewhat unexpected behaviour. The slope of the best fit line for the
distance range of $40\ \le\ d\ \le\ 80\ \Mpch$ exceeds unity, $a=1.22$, \ie there is an excess of
power compared with the target and the Ex/WF cases. This is unexpected for a Bayesian algorithm. The
best linear fit for the range of $120\ \le\ d\ \le\ 160\ \Mpch$ yields a significant negative offset
of $b=-151 \kms$, in agreement with the visual inspection of \cref{fig:vr_sgz}g.

The correlation of the radial component of the velocity field between the reconstructions and the target is shown as a function of
distance in \cref{fig:vr_corr}. Similar to \cref{fig:divv_corr}, these are the correlation coefficients computed from scatter plots
(\cref{fig:scatters_vr}) plotted as a function of distance in order to gauge the degradation of
the reconstruction methods as data becomes more sparse and volumes become large. The solid lines in \cref{fig:vr_corr} represent the mean correlation coefficient between the reconstruction and the target; the error corridor represents the 2$\sigma$ variance about this mean. As expected the Ex/WF is always a
superior to both the BGc/WF and the \Hamlet method. The Ex/WF reconstruction is very well correlated with the target out until $\sim 80\Mpch$, beyond which it begins to drop, although it is worth noting that it stays correlated for the full sample. This drop is a manifestation of the sampling and the decreasing number of the data (per volume) at these distances. The \Hamlet\ and BGc/WF method are roughly equal in the inner regions out to $\sim 70\Mpch$, and beyond it the \Hamlet\ method provides better correlation. At the edge of the data, no method achieves a correlation coefficient of greater than 0.5.

\subsection{Multipole moments of the reconstructed velocity field}

The first two moments of the velocity field, the  monopole and dipole, are examined here. The effect
of errors  and sampling on the fidelity of these two physical quantities is of particular interest
since the monopole and dipole are often used as probes of the scale of homogeneity and can affect
probes of the cosmological model in particular.

\cref{fig:moments}a shows the target and reconstructed velocity monopole as a function of
distance. The same colouring and line style convention used in \cref{fig:divv_corr} and
\cref{fig:vr_corr} is adopted here too, with the moments of the target simulation plotted in
black. Note that the monopole - the mean infall or outflow of matter, is the zeroth order moment of
the velocity field. It is the mean of the divergence of the velocity field
in spheres of radius $d$ and as such is called the ``breathing mode'' of the velocity field. In
    the linear theory of the cosmological gravitational instability the density and velocity fields
    are related by \cref{eq:delta}, hence we opted here to present the
    monopole term by means of this equations.  Thereby,
    \cref{fig:moments}a effectively presents the mean linear density with spheres of radius $d$.
The Ex/WF is nearly indistinguishable from the target here: the error corridor
(which corresponds to variance across all the constrained realisations) is tiny and the black and
green dashed line are practically on top of each other.

The BGc/WF curve overestimates the monopole in the inner parts (within $\sim50\ \Mpch$) while
underestimating it outside that range. This increased monopole implies an overestimation of the
density in the inner parts of the mock universe, which is confirmed by examining the equation of the
best fit line to the scatter plot \cref{fig:scatters_divv} (upper row, left column, $d < 40 \Mpch$).
The best fit line has has an offset of $b=0.11$, meaning that there is a systematic increase in the
estimated densities, consistent with the higher monopole. Both the reconstructions and the target
tend to zero infall at these large scales. The \Hamlet\ method on the other hand behaves inversely
to the BGc/WF method, underestimating the target monopole at small scales and over estimating it at
large scales. The \Hamlet's monopole term at the edge of the data reveals an excess of density at $d
\sim (120 - 150) \Mpch$, in agreement with \cref{fig:scatters_divv} (lower/right panel). Otherwise,
the \Hamlet\ method succeeds in tracking the target monopole over a large range from $\sim 20$ to
$\sim 100 \Mpch$. 

\cref{fig:moments}b and c shows the second moment of the velocity field, namely the dipole or the
bulk flow. \cref{fig:moments}b refers to the magnitude of the bulk flow, while \cref{fig:moments}c
refers to its direction. Accordingly all methods do a fine job of recovering the magnitude of the
bulk flow beyond around $\sim30\Mpch$. The Ex/WF has, predictably, a smaller error corridor than the
other two methods, which are roughly similar in size. With respect to direction, \cref{fig:moments}c
shows the dot product between the target bulk flow direction and the reconstructed one (hence in
this plot there is no black target line). The bulk flow directions for the \Hamlet\ and BGc/WF
method are aligned to within $\sim 15$ deg of the target out to a distance of $\sim 50\Mpch$, while
the Ex/WF is well aligned to greater distances. Note that however, even the Ex/WF curve begins to
deviate significantly at the reconstructed edge. This indicates that even in the best case scenario
of zero errors, sampling at these great distances is a limiting factor in terms of recovering the
direction of the  cosmic dipole. Note that the accuracy recovered here is also restricted in its
ability to recover the underlying dipole direction by the limited  depth of the survey
\citep{Nusser2014}. The problem is exacerbated when examining the BGc/WF and \Hamlet\ curves at
large distances. \cref{fig:moments} indicates that although the monopole and dipole are well
recovered across a large range, the direction of the reconstructed dipole begins to deteriorate when
the sampling drops.

\section{Summary}
\label{sec:conclusion}

The reconstruction of the large scale density and velocity fields from Cosmicflows-like databases
of galaxy distances, and hence peculiar radial velocities, is challenging. The data is sparse,
extremely noisy with Noise/Signal ratio larger than a few for the majority of the data,
non-uniformly and anisotropically  distributed. Furthermore the data suffers from the log-normal
bias, which leads to a non-linear bias in the estimated distances and velocities. 

A number of independent methods have been developed to reconstruct the local LSS and to produce constrained initial
conditions for cosmological simulations designed to reproduce our local patch of the Universe \citep[i.e][]{Sorce2015}. What is generally missing from the literature in this field is an understanding of the accuracy of these methods. Often the reconstructions are applied directly to observational data and only very limited conclusions can be drawn on the viability of a cosmography. The present paper compares the BGc/WF \citep{Hoffman2021} and the \Hamlet\
algorithms \citep{Valade2022} by testing them against a carefully crafted mock of an observational catalogue (an improved CF3-like survey) drawn from one the MultiDark cosmological simulations. 

The quality of the reconstruction is gauged by studying the residual between the reconstructed and
target density and velocity fields. The residual is mostly analyzed by quadratic measures and as
such it is characterized by the mean and variance of the distribution. An optimal reconstruction
should make the mean of the residual to be as close as possible to the null field and aim at
minimizing its variance. A related measure is the linear correlation analysis which yields the
best ``line'', $y = a x + b$, that fits the linear dependence of reconstructed field on the target
one, and the Pearson correlation coefficient. The values of the offset, $b$, for the case of the
linear over-density and for the radial velocity are consistent with zero fro the BGc/WF, in
agreement with the theoretical expectations. The distant data points are extremely noisy and very
sparsely distributed, hence the WF reconstruction is dominated by the \LCDM\ prior model. The
\Hamlet's significant offset is however inconsistent with the prior model.
 
We define here three different regions: the nearby ($d \lesssim 40 \Mpch$), the intermediate (($ 40
\lesssim d  \lesssim 120 \Mpch$) and the distant one ($d \gtrsim 120 \Mpch$).  Based on the above
criteria we conclude that nearby  the BGc/WF and the \Hamlet\ methods are doing roughly equally
well. The methods  diverge at large distance - with the \Hamlet\ outperforming the BGc/WF with a
tighter correlation and smaller variance but underperforming in terms of the bias. This is most
noticeable for  distant region (the right columns of \cref{fig:scatters_divv} and
\cref{fig:divv_corr}).

The three panels of Fig. \ref{fig:moments} deserve a special attention here. The upper panel shows
the radial profile of the monopole moment. The four profiles shown there - target, Ex/WF, BGc/WF and
\Hamlet\ - are all constructed under the assumption of \LCDM\ value of $H_0 = 67.7\,{\rm km/s/Mpc}$. Yet,
the negative offset of the monopole moment at the edge of the data implies that the local value of
$H_0$ is somewhat smaller than its global value. A phenomenon expected for any finite volume
realization in the \LCDM\ cosmology (see \citet{Hoffman2021} for a quantitative assessment). A
proper adjustment of  the local value of $H_0$ would bring the target and Ex/WF profiles to converge
to zero at the edge of the data, together with the BGc/WF asymptotic value. This would leave the
\Hamlet\ positive offset standing out with a systematic bias. The amplitude of the dipole moment,
namely the bulk velocity, is recovered equally well by the three reconstruction and is in very good
agreement with the target. The bottom panel shows the cosine of the angle between the reconstructed
and the target bulk velocities. The BGc/WF behaves as expected - the mean misalignment is consistent
with the full alignment to within one $\sigma$ of the constrained variance. This is not the case
with the \Hamlet\ reconstruction, where the misalignment is more than $2 \sigma$ away from the
expected alignment.  

Our overall assessment of the \Hamlet\ and the  BGc/WF reconstructions is that the former
outperforms the latter one in terms of reduced scatter and tighter correlation between the
reconstructed and the target density and velocity fields. Yet, the \Hamlet\ suffers from biases in
the reconstructed LSS at the distant regime - ones that do not appear in the BGc/WF reconstruction.
It follows that the \Hamlet\ should be the method of choice for the reconstruction of the LSS and
the study of the cosmography of our local patch of the Universe. The BGc/WF reconstruction is the
preferred tool for performing quantitative analysis and parameters estimation and possibly also for
setting initial conditions for constrained cosmological simulations. One last comment is due here.
The WF/CRs is a very well tested approach that is based on a solid theoretical foundations
\citep{Hoffman1992, Zaroubi1995, Zaroubi1999}. As such it provides an attractive framework for
performing Bayesian reconstruction of the nearby LSS. Yet, any bias in the observational data and in
particular the log-normal one  needs to be addressed and apply outside that framework in some ad-hoc
and approximate way. The HMC methodology, and in particular its \Hamlet\ implementation, still
suffer from some teething problems that need to be overcome. The ability of the MCMC methodology in
general and the HMC in particular to address the issue of reconstruction of the LSS, the handling of
observational biases  and the estimation of cosmological parameters within one computational
self-consistent framework makes \Hamlet\ a very attractive tool in the CLUES' toolbox. The
incredible improvement in the computational efficiency of the \Hamlet\ compared with previous
implementation of MCMC algorithms makes it even more promising for future implementations within the
CLUES project.

\section*{Acknowledgements}

Useful discussions with Tamara Davis, concerning the \cite{Hinton2017} paper, are acknowledged. This
work has been done within the framework of the Constrained Local UniversE Simulations (CLUES)
simulations. AV and NIL acknowledge financial support from the Project IDEXLYON at the University of
Lyon under the Investments for the Future Program (ANR-16-IDEX-0005). YH has been partially
supported by the Israel Science Foundation grant ISF 1358/18.

\section*{Data availability}
 	
The data underlying this article will be shared on reasonable request to the corresponding author.

\bibliographystyle{mnras}
\bibliography{bibliography} 

\end{document}